\documentclass{Interspeech2024}

\usepackage{booktabs, multirow} 
\usepackage{soul}
\usepackage{changepage,threeparttable} 
\usepackage[table,xcdraw]{xcolor}
\usepackage{amssymb}
\usepackage{pifont}
\usepackage{siunitx}
\usepackage{caption}
\newcommand{\cmark}{\ding{51}}%
\newcommand{\xmark}{\ding{55}}%

\interspeechcameraready

\title{Low Bitrate High-Quality RVQGAN-based Discrete Speech Tokenizer}

\name{Slava}{Shechtman}
\name{Avihu}{Dekel}

\address{IBM Research}
\email{slava@il.ibm.com, avihu.dekel@ibm.com}

\keywords{Speech Coding, Discrete Speech Tokenization}

\begin{document}

\maketitle

\begin{abstract}
Discrete Audio codecs (or audio tokenizers) have recently regained interest due to the ability of Large Language Models (LLMs) to learn their compressed acoustic representations. Various publicly available trainable discrete tokenizers recently demonstrated impressive results for audio tokenization, yet they mostly require high token rates to gain high-quality reconstruction. In this study, we fine-tuned an open-source general audio RVQGAN model using diverse open-source speech data, considering various recording conditions and quality levels. The resulting wideband (24kHz) speech-only model achieves speech reconstruction, which is nearly indistinguishable from PCM (pulse-code modulation) with a rate of 150-300 tokens per second (1500-3000 bps). The evaluation used comprehensive English speech data encompassing different recording conditions, including studio settings. Speech samples and details on the reproducible trained models are made publicly available\footnote{ 
\url{ibm.biz/IS24SpeechRVQ}}. The model is officially released\footnote{ 
\url{https://huggingface.co/ibm/DAC.speech.v1.0}}

\end{abstract}

\section{Introduction}
\label{sec:intro}

Digital speech and audio codecs were originally used solely for efficient audio compression for data transmission~\cite{OPUS, EVS}. 
In such settings, the encoder transforms the digitally sampled audio signal (with a sampling rate that can vary from Narrow Band 8kHz for telephony to Full Band 48-96kHz for high-fidelity studio recordings) into a compressed digital stream that is subsequently transmitted. 
Before encoding, audio signals typically undergo a two-step process. First, {\em segmentation} divides the audio data into smaller units known as frames. Next, {\em feature extraction} represents these frames in a manner that enables efficient lossy signal compression while minimizing perceptual degradation for human listeners.
Certain features, such as Spectrograms or Mel-Frequency Cepstral Coefficients, characterize the spectral envelope of the signal frame, while the remaining ones describe the time-domain residual signal.
In classic codecs, the spectral features usually undergo multiple trainable Vector Quantizations (VQ), while the rest are represented with a mixture of trainable and rule-based codebooks along with binary representations of various control parameters~\cite{OPUS, EVS}. 
Once transmitted,
the receiver employs the decoder to reconstruct the original audio from the received digital stream. 

\begin{figure}
    \centering
    \includegraphics[width=\linewidth]{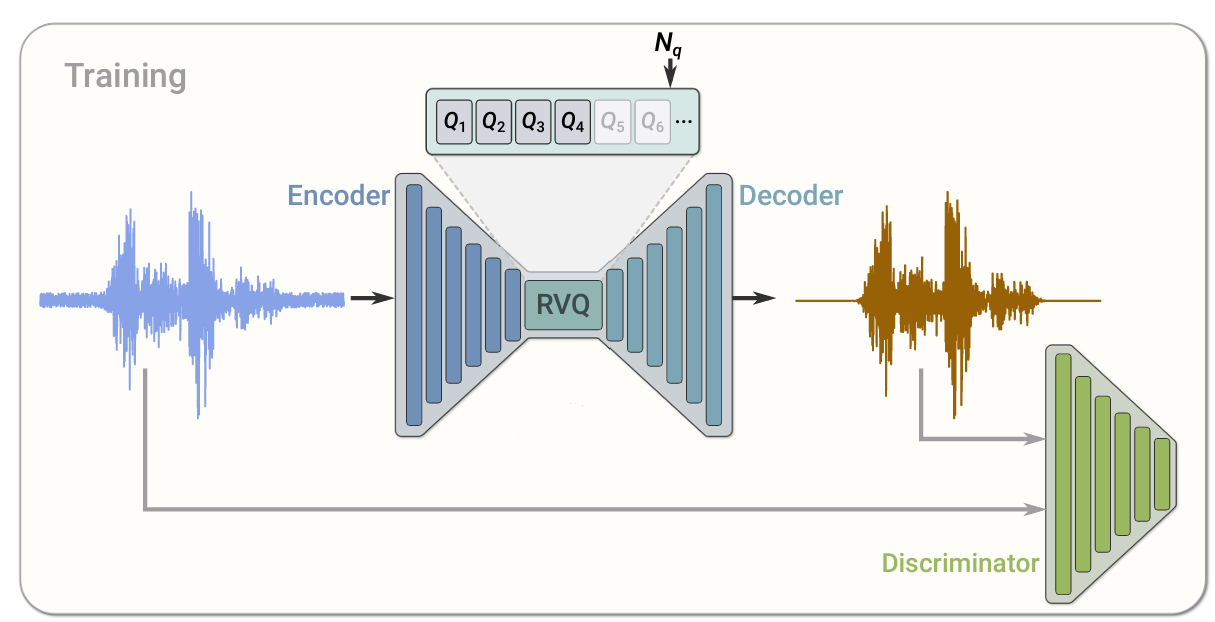}
    \caption{VQGAN architecture overview~\cite{Soundstream}}
    \label{fig:RVQ}
\end{figure}

Discrete audio tokenization, a special case of digital audio coding, has recently gained renewed interest due to its potential for applying Large Language Modeling (LLM) techniques in audio and mixed text/audio domains for one-shot speech synthesis~\cite{VALLE, AudioLM, SoundStorm}, speech recognition~\cite{2023viola}, speaker recognition ~\cite{puvvada2023discrete} and more. 
Unlike classic audio coding, which imposes no constraints on the digital-encoder output, discrete audio tokenization~\cite{Soundstream} (see Figure~\ref{fig:RVQ}) involves converting an audio signal into a sequence of discrete tokens, represented by integers serving as codeword indices associated with a set of trained codebooks. 
Audio tokens are designed to capture essential details for precise audio reconstruction using a trainable decoder. This enables the audio tokens generated by a language model to be readily converted into high-quality audio waveforms.

Discrete tokenizers usually feature a classic bottleneck autoencoder structure with a quantization layer~\cite{VQVAE}. The Residual Vector Quantization with Generative Adversary Network (RVQGAN) based architecture~\cite{Soundstream}  stands out as one of the most popular choices for tokenization. It deploys residual vector quantization (RVQ)~\cite{barnes1996RVQAdvancesReview} of the bottleneck features, where each quantization layer refines the previous quantization layer, resulting in multiple tokens representing a single audio frame. 
Several discrete tokenizers, suitable for speech or general audio, have recently become available and gained popularity in the open-source community~\cite{Encodec, DAC, Funcodec}. 
They are mostly RVQGAN-based and their high-quality operating points start from 600 tokens per second~\cite{wu2024towards},
a demanding high data rate to efficiently train LLM models. To improve the discrete generative modeling of audio, it is highly beneficial to compress further the sequence of tokens representing an audio sample, while preserving the high fidelity of the audio reconstruction.

We base this work on the Descript Audio Codec (DAC)~\cite{DAC},
an open-source universal discrete tokenizer model capable of preserving high-fidelity audio quality across diverse audio material (including general audio, music, and speech) at bit rates of 6-8 kbps. 
However, its performance dramatically deteriorates for bitrates of 3 kbps and below~\cite{DAC}. 

The goal of the current work is to adapt the universal audio DAC model to a high-quality speech-only discrete tokenizer model with reduced operational bitrates.
Our contributions are summed up as follows:
\begin{itemize}
\item We fine-tune a universal DAC model in low-bitrate settings 
(1500-3000 bps) with diverse open-source speech data, while carefully balancing various recording conditions and audio quality levels, and focusing on high-fidelity speech data. 
\item We evaluate the resulting models on various speech datasets, demonstrating high-quality reconstruction for the 1.5-kbps model, and perceptually transparent reconstruction for the 3-kbps model.
\item We perform a thorough ablation study to explore how training speech data of various quality levels and recording conditions influences the model performance, as assessed on versatile test data.      
\end{itemize}


\section{Method}
\label{sec:model}
\subsection{RVQ-GAN Model}
Residual Vector Quantization (RVQ) is a classic speech-coding technique~\cite{vasuki2006review} that has been recently revived as a key element in modern neural discrete tokenization when combined with Generative Adversarial Network (GAN)~\cite{GAN} training techniques~\cite{Soundstream}. 
An overview of the ensuing RVQGAN model is presented in Figure~\ref{fig:RVQ}. 
In this autoencoder architecture, an encoder downsamples an input signal, creating a more compact latent representation that is incrementally quantized by RVQ, and then reconstructed by a decoder whose structure mirrors that of the encoder. RVQ is a multi-stage VQ technique, where each stage quantizes the residual from the previous VQ stage. 
During training, the model parameters are optimized using a combination of reconstruction and adversarial losses, where a separate discriminator network is trained concurrently with the autoencoder network~\cite{Soundstream}.
The non-differential quantization layers are optimized using the straight-through estimator \cite{VQVAE}.


The Descript Audio Codec (DAC)~\cite{DAC} is an RVQGAN-based universal audio codec model that is remarkable in its capability to preserve high-fidelity audio quality across diverse audio material, including general audio, music, and speech~\cite{DAC}. It achieves this by incorporating several techniques, such as periodic activation functions, codebook factorization with L2-normalization, and improved reconstruction and adversarial losses~\cite{DAC}. It also deploys random quantizer dropout to support multiple bitrates for a single model and stabilize the training~\cite{Soundstream}. Both pretrained DAC models and model source code are released as open-source\footnote{\href{https://github.com/descriptinc/descript-audio-codec}{ https://github.com/descriptinc/descript-audio-codec}}. They also have been shown to outperform several former popular discrete tokenization models \cite{Soundstream,Encodec}.


\begin{figure}
    \centering
    \caption{MUSHRA results with 95\% confidence interval}
\includegraphics[width=\linewidth]{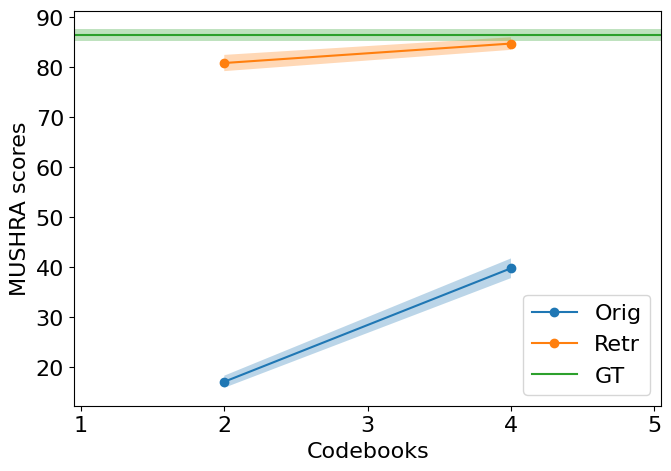}
    \label{fig:mushra}
\end{figure}


\subsection{Training Data Selection}
\label{sec:data}

The quality of discrete audio tokenizers depends not only on their architecture and training protocols but also on the quality of the training data,
and among those datasets commonly used for training a discrete tokenizer~\cite{wu2024towards}, we observe data of variable speech quality.
In general, trainable tokenizers are not language-dependent provided multi-lingual data is used for training. However, in this work, we only focus on English, so we refer below just to English datasets (or English sections of multi-lingual ones), serving for DAC training~\cite{DAC}: 

\begin{table}[ht]
    \caption{Wilcoxon Rank Sum Test comparing systems to the raw audio}
    \label{table:wilcoxon}
    \centering
    \begin{tabular}{c c c c c }
    \toprule
       System  & Orig-2 & Orig-4 & Retr-2 & Retr-4 \\
       \midrule
       p-value & \num{<1e-5} & \num{<1e-5} & \num{<1e-5} & \num{0.062} \\
     \bottomrule
    \end{tabular}
\end{table}

\begin{itemize}
\item \emph{ReadSpeech} is a high-quality full-band (48kHz) dataset from the Denoising Challenge \cite{DNS}, containing about 1000 hrs. It is mostly derived from audiobooks with good recording conditions.
\item \emph{DAPS}~\cite{DAPS} is a small (4.5hrs) dataset of studio quality full band (48khz) recordings.
\item \emph{LibriSpeech}~\cite{LibriSpeech} is a large (around 1000 hrs) 16kHz-sampled dataset of medium quality that is also based on public-domain audiobooks.
\item \emph{LibriTTS}~\cite{LibriTTS} (585 hrs) is also derived from audiobooks, and it is a source of full band (24kHz) speech of medium quality containing a \emph{LibriTTS-clean} subset with better recording conditions, and a \emph{LibriTTS-other} subset with heavily accented speech and/or more challenging recording setup (including noise and reverberation). It is derived from the same material as the LibriSpeech dataset.
\item \emph{VCTK}~\cite{VCTK} is a medium quality full-band (48kHz) speech dataset (44hrs, 110 speakers).
\item \emph{Common Voice}~\cite{CommonVoice} (2500 hrs) is a crowd-collected dataset of speech sampled at 8-16kHz, featuring the most challenging recording conditions and distortions.
\end{itemize}

For the task of the DAC model adaptation for speech-only settings (see Section~\ref{sub:impl}), we added extra sources of speech data: 
\begin{itemize}
\item \emph{LJ-Speech}~\cite{ljspeech17}, a 24-hrs long full-band single-speaker dataset of studio quality.
\item \emph{LibriTTS-R}~\cite{koizumilibrittsR}, a high fidelity version of \emph{LibriTTS} that is created by passing the whole \emph{LibriTTS} dataset through a Speech Restoration network, simulating clean studio recording conditions. 
\item \emph{LibriLight}~\cite{librilight} very large (60k hours) mixed-quality public domain dataset of 16khz speech. In our retraining, We used this dataset as an unqualified data source, instead of \emph{Common Voice}.
\end{itemize}

The training data does not equally represent the various quality levels. To address
this, the original training procedure employs balanced data sampling, ensuring an equal mix of datasets from different sources and quality levels within each training mini-batch~\cite{DAC}. 
We adopted this approach for the model adaptation, too, and divided our training data into the following categories, which are equally balanced over the training mini-batches:
\begin{itemize}
\item \emph{HQ1}: high-quality, clean, contains \emph{ReadSpeech}, \emph{DAPS}, and \emph{LJ-speech} datasets.
\item \emph{HQ2}: restored high-quality, clean, contains the \emph{LibriTTS-R-clean} portion of \emph{LibriTTS-R}.
\item \emph{HQ3}: restored high-quality, clean, contains the \emph{LibriTTS-R-other} portion of \emph{LibriTTS-R}, where more challenging accentuation is present.
\item \emph{MQ1}: medium-quality, clean, contains the \emph{LibriTTS-clean} portion of \emph{LibriTTS}.
\item \emph{MQ2}: medium-quality, unclean, contains the \emph{LibriTTS-other} portion of \emph{LibriTTS}.
\item \emph{UQ}: unqualified (low-quality/mixed quality), unclean, contains \emph{LibriLight} dataset, upsampled to 24kHz
\end{itemize} 
One can notice that in the proposed setup the low-quality data is under-represented. We found these settings beneficial for high-quality speech reconstruction. Additional data selection trends are explored in Sec~\ref{sub:dataabl}       

\subsection{Implementation Details}
\label{sub:impl}
We retrained a 24-kHz universal audio DAC model by strictly following the training procedure proposed in~\cite{DAC} with the balanced training data as detailed above. Unlike the original DAC training, we omitted quantizer dropout (the procedure of random dropping out of some of the later stages in RVQ, during the training~\cite{Soundstream}). While the quantizer dropout proved beneficial when training from scratch~\cite{DAC}, we observed that it had a detrimental effect on model performance when adapting from a pretrained model. 

We trained a set of fixed bitrate models (with no quantization dropout) each utilizing a different number of 10-bit RVQ codebooks  $Q\in \{1,2,4,8,16,32\}$ (corresponding to bitrates $R\in \{0.375,0.75,3,6,12,24\}$ kbps), initialized from the publicly available 24kbps DAC model for 24kHz audio~\cite{DAC}. The original model encodes an audio frame with 1-32 RVQ codebooks of 10 bits each at a frame rate of 75Hz. Training took place on two \emph{A100\_80g} GPUs for 400k steps with a mini-batch of 72 excerpts of a fixed length of 0.38 sec (randomly extracted from longer speech samples in the datasets).   


\begin{table}[ht]
\centering
\caption{Objective metrics for Retrained speech model (\emph{Retr}) vs. Original model (\emph{Orig}) as assessed on various test sets}\label{table:obj_res}
\scriptsize
\begin{tabular}{lrrrrrrrrr}\toprule
\multirow{2}{*}{test data} &\multirow{2}{*}{metric} &\multirow{2}{*}{system} &\multicolumn{6}{c}{number of codebooks} \\\cmidrule{4-9}
& & &32 &16 &8 &4 &2 &1 \\\midrule
\multirow{8}{*}{studio} &\multirow{2}{*}{mel↓} &Orig &\cellcolor[HTML]{93cc7d}0.22 &\cellcolor[HTML]{c4d980}0.29 &\cellcolor[HTML]{ffe984}0.38 &\cellcolor[HTML]{fdc07c}0.48 &\cellcolor[HTML]{fb9774}0.58 &\cellcolor[HTML]{f8696b}0.69 \\
& &Retr &\cellcolor[HTML]{63be7b}\textbf{0.15} &\cellcolor[HTML]{93cc7d}\textbf{0.22} &\cellcolor[HTML]{cbdc81}\textbf{0.3} &\cellcolor[HTML]{fbea83}\textbf{0.37} &\cellcolor[HTML]{fecd7f}\textbf{0.45} &\cellcolor[HTML]{fb9b75}\textbf{0.57} \\
&\multirow{2}{*}{STFT↓} &Orig &\cellcolor[HTML]{8ac97d}0.47 &\cellcolor[HTML]{cbdc81}0.55 &\cellcolor[HTML]{ffe383}0.64 &\cellcolor[HTML]{fdc07c}0.75 &\cellcolor[HTML]{fb9073}0.9 &\cellcolor[HTML]{f8696b}1.02 \\
& &Retr &\cellcolor[HTML]{63be7b}\textbf{0.42} &\cellcolor[HTML]{9bce7e}\textbf{0.49} &\cellcolor[HTML]{bbd780}\textbf{0.53} &\cellcolor[HTML]{eae582}\textbf{0.59} &\cellcolor[HTML]{ffda81}\textbf{0.67} &\cellcolor[HTML]{fcb37a}\textbf{0.79} \\
&\multirow{2}{*}{PESQ↑} &Orig &\cellcolor[HTML]{6ac07c}4.5 &\cellcolor[HTML]{8dca7e}4.3 &\cellcolor[HTML]{f5e984}3.69 &\cellcolor[HTML]{fbb279}2.6 &\cellcolor[HTML]{f9806f}1.67 &\cellcolor[HTML]{f8696b}1.24 \\
& &Retr &\cellcolor[HTML]{63be7b}\textbf{4.54} &\cellcolor[HTML]{76c47d}\textbf{4.43} &\cellcolor[HTML]{add480}\textbf{4.11} &\cellcolor[HTML]{fee783}\textbf{3.57} &\cellcolor[HTML]{fcbc7b}\textbf{2.78} &\cellcolor[HTML]{f98d72}\textbf{1.91} \\
&\multirow{2}{*}{STOI↑} &Orig &\cellcolor[HTML]{63be7b}\textbf{1} &\cellcolor[HTML]{98ce7f}\textbf{0.99} &\cellcolor[HTML]{ffeb84}0.97 &\cellcolor[HTML]{fdce7e}0.93 &\cellcolor[HTML]{fba276}0.87 &\cellcolor[HTML]{f8696b}0.79 \\
& &Retr &\cellcolor[HTML]{63be7b}\textbf{1} &\cellcolor[HTML]{98ce7f}\textbf{0.99} &\cellcolor[HTML]{ccdd82}\textbf{0.98} &\cellcolor[HTML]{ffeb84}\textbf{0.97} &\cellcolor[HTML]{fedc81}\textbf{0.95} &\cellcolor[HTML]{fbb178}\textbf{0.89} \\\midrule
& &Orig & \cellcolor[HTML]{8CC97D}0.56 & \cellcolor[HTML]{BED880}0.75    & \cellcolor[HTML]{FCEA83}0.98  & \cellcolor[HTML]{FEC77D}1.22          & \cellcolor[HTML]{FB9E76}1.47  & \cellcolor[HTML]{F8696B}1.79 \\
& \multirow{-2}{*}{mel↓}  &Retr &\cellcolor[HTML]{63BE7B}\textbf{0.41} 
& \cellcolor[HTML]{98CD7E}\textbf{0.61} & \cellcolor[HTML]{CFDD81}\textbf{0.82} & \cellcolor[HTML]{FFEA84}\textbf{1.00} & \cellcolor[HTML]{FEC97E}\textbf{1.20} & \cellcolor[HTML]{FBA176}\textbf{1.45} \\
& &Orig & \cellcolor[HTML]{6CC07B}1.15 & \cellcolor[HTML]{AAD27F}1.31    & \cellcolor[HTML]{F4E883}1.51 & \cellcolor[HTML]{FEC97E}1.71 
& \cellcolor[HTML]{FB9F76}1.91          & \cellcolor[HTML]{F8696B}2.18 \\
& \multirow{-2}{*}{STFT↓}  & Retr                     & \cellcolor[HTML]{63BE7B}\textbf{1.12} & \cellcolor[HTML]{9DCE7E}\textbf{1.28} & \cellcolor[HTML]{D5DE81}\textbf{1.43} & \cellcolor[HTML]{FFE683}\textbf{1.56} & \cellcolor[HTML]{FDC57D}\textbf{1.72} & \cellcolor[HTML]{FB9E76}\textbf{1.92} \\
&                          & Orig                     & \cellcolor[HTML]{68C07C}4.50          & \cellcolor[HTML]{86C87D}4.32          & \cellcolor[HTML]{F5E884}3.65          & \cellcolor[HTML]{FCB579}2.63          & \cellcolor[HTML]{F98370}1.74          & \cellcolor[HTML]{F8696B}1.28          \\
& \multirow{-2}{*}{PESQ↑}  & Retr                     & \cellcolor[HTML]{63BE7B}\textbf{4.52} & \cellcolor[HTML]{78C47D}\textbf{4.40} & \cellcolor[HTML]{B1D580}\textbf{4.06} & \cellcolor[HTML]{FEE783}\textbf{3.52} & \cellcolor[HTML]{FCBC7B}\textbf{2.76} & \cellcolor[HTML]{FA9072}\textbf{1.98} \\
&                          & Orig                     & \cellcolor[HTML]{67C07C}1.00          & \cellcolor[HTML]{98CE7F}0.98          & \cellcolor[HTML]{FFEB84}0.95          & \cellcolor[HTML]{FDCC7E}0.91          & \cellcolor[HTML]{FBA175}0.86          & \cellcolor[HTML]{F8696B}0.78          \\
\multirow{-8}{*}{DAPS}              & \multirow{-2}{*}{STOI↑}  & Retr                     & \cellcolor[HTML]{63BE7B}\textbf{1.00} & \cellcolor[HTML]{81C77D}\textbf{0.99} & \cellcolor[HTML]{B5D680}\textbf{0.97} & \cellcolor[HTML]{FFEB84}\textbf{0.95} & \cellcolor[HTML]{FDD37F}\textbf{0.92} & \cellcolor[HTML]{FBAB77}\textbf{0.87} \\\midrule          

\multirow{8}{2em}{Libri-TTS-R-clean} &\multirow{2}{*}{mel ↓} &Orig &\cellcolor[HTML]{96cc7d}0.25 &\cellcolor[HTML]{cbdc81}0.33 &\cellcolor[HTML]{ffe583}0.43 &\cellcolor[HTML]{fdc27d}0.53 &\cellcolor[HTML]{fb9975}0.65 &\cellcolor[HTML]{f8696b}0.79 \\
& &Retr &\cellcolor[HTML]{63be7b}\textbf{0.17} &\cellcolor[HTML]{96cc7d}\textbf{0.25} &\cellcolor[HTML]{c4da80}\textbf{0.32} &\cellcolor[HTML]{f2e783}\textbf{0.39} &\cellcolor[HTML]{fed781}\textbf{0.47} &\cellcolor[HTML]{fcb179}\textbf{0.58} \\
&\multirow{2}{*}{STFT↓} &Orig &\cellcolor[HTML]{97cd7e}0.5 &\cellcolor[HTML]{d3de81}0.57 &\cellcolor[HTML]{ffdc82}0.66 &\cellcolor[HTML]{fdba7b}0.75 &\cellcolor[HTML]{fb9474}0.85 &\cellcolor[HTML]{f8696b}0.96 \\
& &Retr &\cellcolor[HTML]{63be7b}\textbf{0.44} &\cellcolor[HTML]{97cd7e}\textbf{0.5} &\cellcolor[HTML]{c2d980}\textbf{0.55} &\cellcolor[HTML]{e4e382}\textbf{0.59} &\cellcolor[HTML]{ffe082}\textbf{0.65} &\cellcolor[HTML]{fdbe7c}\textbf{0.74} \\
&\multirow{2}{*}{PESQ↑} &Orig &\cellcolor[HTML]{6bc17c}4.5 &\cellcolor[HTML]{90cb7e}4.3 &\cellcolor[HTML]{fee983}3.67 &\cellcolor[HTML]{fbb178}2.62 &\cellcolor[HTML]{f98370}1.75 &\cellcolor[HTML]{f8696b}1.25 \\
& &Retr &\cellcolor[HTML]{63be7b}\textbf{4.54} &\cellcolor[HTML]{76c47d}\textbf{4.44} &\cellcolor[HTML]{acd480}\textbf{4.15} &\cellcolor[HTML]{f9ea84}\textbf{3.74} &\cellcolor[HTML]{fdc87d}\textbf{3.06} &\cellcolor[HTML]{fa9974}\textbf{2.17} \\
&\multirow{2}{*}{STOI↑} &Orig &\cellcolor[HTML]{63be7b}\textbf{1} &\cellcolor[HTML]{98ce7f}\textbf{0.99} &\cellcolor[HTML]{ffeb84}0.97 &\cellcolor[HTML]{fdca7d}0.93 &\cellcolor[HTML]{fba175}0.88 &\cellcolor[HTML]{f8696b}0.81 \\
& &Retr &\cellcolor[HTML]{63be7b}\textbf{1} &\cellcolor[HTML]{98ce7f}\textbf{0.99} &\cellcolor[HTML]{ccdd82}\textbf{0.98} &\cellcolor[HTML]{ffeb84}\textbf{0.97} &\cellcolor[HTML]{feda80}\textbf{0.95} &\cellcolor[HTML]{fcba7a}\textbf{0.91} \\\midrule

& \multirow{-2}{*}{mel ↓}  & Retr                     & \cellcolor[HTML]{63BE7B}\textbf{0.13} & \cellcolor[HTML]{96CC7D}\textbf{0.19} & \cellcolor[HTML]{CADB80}\textbf{0.25} & \cellcolor[HTML]{F3E783}\textbf{0.30} & \cellcolor[HTML]{FED781}\textbf{0.35} & \cellcolor[HTML]{FCB279}\textbf{0.44} \\
&                          & Orig                     & \cellcolor[HTML]{93CB7D}0.38          & \cellcolor[HTML]{D1DD81}0.44          & \cellcolor[HTML]{FFE082}0.50          & \cellcolor[HTML]{FDBF7C}0.56          & \cellcolor[HTML]{FB9875}0.63          & \cellcolor[HTML]{F8696B}0.72          \\
& \multirow{-2}{*}{STFT↓}  & Retr                     & \cellcolor[HTML]{63BE7B}\textbf{0.34} & \cellcolor[HTML]{97CD7E}\textbf{0.39} & \cellcolor[HTML]{C1D980}\textbf{0.42} & \cellcolor[HTML]{E5E382}\textbf{0.45} & \cellcolor[HTML]{FFE082}\textbf{0.50} & \cellcolor[HTML]{FDBF7C}\textbf{0.56} \\
&                          & Orig                     & \cellcolor[HTML]{65BF7C}4.50          & \cellcolor[HTML]{6EC17C}4.31          & \cellcolor[HTML]{89C97E}3.68          & \cellcolor[HTML]{B3D680}2.69          & \cellcolor[HTML]{D8E082}1.83          & \cellcolor[HTML]{EFE784}1.30          \\
& \multirow{-2}{*}{PESQ↑}  & Retr                     & \cellcolor[HTML]{63BE7B}\textbf{4.55} & \cellcolor[HTML]{68C07C}\textbf{4.44} & \cellcolor[HTML]{75C47D}\textbf{4.15} & \cellcolor[HTML]{87C97E}\textbf{3.73} & \cellcolor[HTML]{A4D17F}\textbf{3.06} & \cellcolor[HTML]{C9DC81}\textbf{2.19} \\
&                          & Orig                     & \cellcolor[HTML]{69C07C}0.99          & \cellcolor[HTML]{9ECF7F}0.98          & \cellcolor[HTML]{FEE883}0.96          & \cellcolor[HTML]{FDCC7E}0.93          & \cellcolor[HTML]{FBA476}0.88          & \cellcolor[HTML]{F8696B}0.80          \\
\multirow{-8}{2em}{Libri-TTS-R-other} & \multirow{-2}{*}{STOI↑}  & Retr                     & \cellcolor[HTML]{63BE7B}\textbf{1.00} & \cellcolor[HTML]{7EC67D}\textbf{0.99} & \cellcolor[HTML]{AED480}\textbf{0.98} & \cellcolor[HTML]{EDE683}\textbf{0.97} & \cellcolor[HTML]{FEDC81}\textbf{0.95} & \cellcolor[HTML]{FCB77A}\textbf{0.90} \\
&                          & Orig                     & \cellcolor[HTML]{93CB7D}0.27          & \cellcolor[HTML]{C2D980}0.35          & \cellcolor[HTML]{FFEB84}0.45          & \cellcolor[HTML]{FDC47D}0.55          & \cellcolor[HTML]{FB9975}0.66          & \cellcolor[HTML]{F8696B}0.78          \\\midrule

\multirow{8}{2em}{Libri-TTS-clean} &\multirow{2}{*}{mel ↓} &Orig &\cellcolor[HTML]{93cb7d}0.27 &\cellcolor[HTML]{c2d980}0.35 &\cellcolor[HTML]{ffeb84}0.45 &\cellcolor[HTML]{fdc47d}0.55 &\cellcolor[HTML]{fb9975}0.66 &\cellcolor[HTML]{f8696b}0.78 \\
& &Retr &\cellcolor[HTML]{63be7b}\textbf{0.19} &\cellcolor[HTML]{99cd7e}\textbf{0.28} &\cellcolor[HTML]{cfdd81}\textbf{0.37} &\cellcolor[HTML]{ffeb84}\textbf{0.45} &\cellcolor[HTML]{fecc7e}\textbf{0.53} &\cellcolor[HTML]{fba176}\textbf{0.64} \\
&\multirow{2}{*}{STFT↓} &Orig &\cellcolor[HTML]{7bc57c}0.53 &\cellcolor[HTML]{bdd880}0.61 &\cellcolor[HTML]{ffe784}0.7 &\cellcolor[HTML]{fdc07c}0.79 &\cellcolor[HTML]{fb9975}0.88 &\cellcolor[HTML]{f8696b}0.99 \\
& &Retr &\cellcolor[HTML]{63be7b}\textbf{0.5} &\cellcolor[HTML]{94cc7d}\textbf{0.56} &\cellcolor[HTML]{cddc81}\textbf{0.63} &\cellcolor[HTML]{f6e883}\textbf{0.68} &\cellcolor[HTML]{fed680}\textbf{0.74} &\cellcolor[HTML]{fcaf79}\textbf{0.83} \\
&\multirow{2}{*}{PESQ↑} &Orig &\cellcolor[HTML]{6bc17c}\textbf{4.46} &\cellcolor[HTML]{8cca7e}\textbf{4.25} &\cellcolor[HTML]{f6e984}3.58 &\cellcolor[HTML]{fbb279}2.52 &\cellcolor[HTML]{f9806f}1.64 &\cellcolor[HTML]{f8696b}1.23 \\
& &Retr &\cellcolor[HTML]{63be7b}\textbf{4.51} &\cellcolor[HTML]{78c47d}\textbf{4.38} &\cellcolor[HTML]{b4d680}\textbf{4} &\cellcolor[HTML]{fee783}\textbf{3.46} &\cellcolor[HTML]{fcba7a}\textbf{2.67} &\cellcolor[HTML]{f98d72}\textbf{1.88} \\
&\multirow{2}{*}{STOI↑} &Orig &\cellcolor[HTML]{63be7b}\textbf{0.99} &\cellcolor[HTML]{98ce7f}\textbf{0.98} &\cellcolor[HTML]{ffeb84}0.96 &\cellcolor[HTML]{fdce7e}0.92 &\cellcolor[HTML]{fba276}0.86 &\cellcolor[HTML]{f8696b}0.78 \\
& &Retr &\cellcolor[HTML]{63be7b}\textbf{0.99} &\cellcolor[HTML]{63be7b}\textbf{0.99} &\cellcolor[HTML]{98ce7f}\textbf{0.98} &\cellcolor[HTML]{ffeb84}\textbf{0.96} &\cellcolor[HTML]{fdd57f}\textbf{0.93} &\cellcolor[HTML]{fbaa77}\textbf{0.87} \\\midrule
\multirow{8}{2em}{Libri-TTS-other} &\multirow{2}{*}{mel ↓} &Orig &\cellcolor[HTML]{90cb7d}0.21 &\cellcolor[HTML]{c5da80}0.28 &\cellcolor[HTML]{fbe983}0.35 &\cellcolor[HTML]{fec77d}0.43 &\cellcolor[HTML]{fb9f76}0.51 &\cellcolor[HTML]{f8696b}0.62 \\
& &Retr &\cellcolor[HTML]{63be7b}\textbf{0.15} &\cellcolor[HTML]{98cd7e}\textbf{0.22} &\cellcolor[HTML]{d5de81}\textbf{0.3} &\cellcolor[HTML]{ffe984}\textbf{0.36} &\cellcolor[HTML]{fec77d}\textbf{0.43} &\cellcolor[HTML]{fb9b75}\textbf{0.52} \\
&\multirow{2}{*}{STFT↓} &Orig &\cellcolor[HTML]{80c67c}0.42 &\cellcolor[HTML]{c4da80}0.49 &\cellcolor[HTML]{ffe683}0.56 &\cellcolor[HTML]{fdc67d}0.62 &\cellcolor[HTML]{fb9a75}0.7 &\cellcolor[HTML]{f8696b}0.79 \\
& &Retr &\cellcolor[HTML]{63be7b}\textbf{0.39} &\cellcolor[HTML]{9dce7e}\textbf{0.45} &\cellcolor[HTML]{cedc81}\textbf{0.5} &\cellcolor[HTML]{f5e883}\textbf{0.54} &\cellcolor[HTML]{fed07f}\textbf{0.6} &\cellcolor[HTML]{fcaa78}\textbf{0.67} \\
&\multirow{2}{*}{PESQ↑} &Orig &\cellcolor[HTML]{6bc17c}\textbf{4.42} &\cellcolor[HTML]{90cb7e}\textbf{4.14} &\cellcolor[HTML]{f1e784}3.39 &\cellcolor[HTML]{fbb379}2.39 &\cellcolor[HTML]{f9816f}1.61 &\cellcolor[HTML]{f8696b}1.22 \\
& &Retr &\cellcolor[HTML]{63be7b}\textbf{4.48} &\cellcolor[HTML]{7bc57d}\textbf{4.3} &\cellcolor[HTML]{bad881}\textbf{3.81} &\cellcolor[HTML]{fee382}\textbf{3.16} &\cellcolor[HTML]{fcb379}\textbf{2.4} &\cellcolor[HTML]{f98971}\textbf{1.73} \\
&\multirow{2}{*}{STOI↑} &Orig &\cellcolor[HTML]{63be7b}\textbf{0.99} &\cellcolor[HTML]{a2d07f}\textbf{0.97} &\cellcolor[HTML]{ffeb84}0.94 &\cellcolor[HTML]{fdcf7e}0.9 &\cellcolor[HTML]{fa9f75}0.83 &\cellcolor[HTML]{f8696b}0.75 \\
& &Retr &\cellcolor[HTML]{63be7b}\textbf{0.99} &\cellcolor[HTML]{83c77d}\textbf{0.98} &\cellcolor[HTML]{c1da81}\textbf{0.96} &\cellcolor[HTML]{ffeb84}\textbf{0.94} &\cellcolor[HTML]{fdcf7e}\textbf{0.9} &\cellcolor[HTML]{fba676}\textbf{0.84} \\
\bottomrule
\end{tabular}
\end{table}

\subsection{Evaluation metrics}
We make use of the following metrics for objective reconstruction evaluations:
\begin{itemize}
\item \emph{mel loss}: a combined mel-scale loss, serving as a mel reconstruction loss during DAC training~\cite{DAC}. It is evaluated as a sum of $L1$-distances (between the ground truth and the reconstructed) of log mel spectrograms of various spectral resolutions.
\item \emph{STFT loss}: a combined Short-Time Fourier Transform (STFT) loss, serving as a linear frequency-domain loss during DAC training~\cite{DAC}. It is evaluated as a sum of $L1$-distances (between the ground truth and reconstructions) of linear spectrograms at various spectral resolutions.
\item \emph{PESQ}: Wideband (16kHz) speech quality assessment score~\cite{PESQ}.
\item \emph{STOI}: speech intelligibility metric~\cite{STOI}.
\end{itemize}

While those metrics serve the purpose of tracking trends and model comparisons, they do not reliably predict perceptually significant distortion. 
Those metrics are therefore complemented with subjective listening evaluations (see Sec~\ref{subsub:eval}).


\begin{table*}[!htp]
\captionsetup{justification=centering}
\centering
\caption{Removing high-quality (HQ) and mid/unknown-quality (MQ/UQ) data sources, \\and measuring objective metrics various speech test sets}
\label{tab:data_ablation}
\scriptsize
\begin{tabular}{llr|rrr|rrrrrr}\toprule
\multicolumn{2}{c|}{Test Data} &full &\multicolumn{3}{c|}{Reduce HQ Data} &\multicolumn{5}{c}{Reduce MQ/UQ Data} \\\midrule
\multicolumn{2}{l|}{LJ-speech} &\cmark &\xmark &\cmark &\xmark &\cmark &\cmark &\cmark &\cmark &\cmark \\
\multicolumn{2}{l|}{Libri-TTS-R-clean} &\cmark &\cmark &\xmark &\xmark &\cmark &\cmark &\cmark &\cmark &\cmark \\
\multicolumn{2}{l|}{Libri-TTS-R-others} &\cmark &\cmark &\xmark &\xmark &\cmark &\cmark &\cmark &\xmark &\xmark \\
\multicolumn{2}{l|}{Libri-TTS-clean} &\cmark &\cmark &\cmark &\cmark &\cmark &\cmark &\xmark &\cmark &\xmark \\
\multicolumn{2}{l|}{Libri-TTS-others} &\cmark &\cmark &\cmark &\cmark &\cmark &\xmark &\xmark &\xmark &\xmark \\
\multicolumn{2}{l|}{UQ} &\cmark &\cmark &\cmark &\cmark &\xmark &\xmark &\xmark &\xmark &\xmark \\\midrule
\multirow{3}{*}{Studio} &mel loss↓ &\cellcolor[HTML]{63be7b}\textbf{0.37} &\cellcolor[HTML]{e2e282}0.38 &\cellcolor[HTML]{ffeb84}0.38 &\cellcolor[HTML]{f0e683}0.38 &\cellcolor[HTML]{ffdf82}0.38 &\cellcolor[HTML]{ffdf82}0.38 &\cellcolor[HTML]{f8696b}0.40 &\cellcolor[HTML]{d4de81}0.38 &\cellcolor[HTML]{f8696b}0.40 \\
&STFT loss↓ &\cellcolor[HTML]{63be7b}\textbf{0.59} &\cellcolor[HTML]{ffe583}0.62 &\cellcolor[HTML]{7dc57c}0.59 &\cellcolor[HTML]{fed280}0.64 &\cellcolor[HTML]{ffeb84}0.62 &\cellcolor[HTML]{ffeb84}0.62 &\cellcolor[HTML]{f8696b}0.71 &\cellcolor[HTML]{abd37f}0.60 &\cellcolor[HTML]{fa8972}0.69 \\
&PESQ↑ &\cellcolor[HTML]{6dc17c}3.57 &\cellcolor[HTML]{63be7b}\textbf{3.58} &\cellcolor[HTML]{fcbb7a}3.48 &\cellcolor[HTML]{dae182}3.53 &\cellcolor[HTML]{ffeb84}3.51 &\cellcolor[HTML]{fedd81}3.50 &\cellcolor[HTML]{f8756d}3.43 &\cellcolor[HTML]{fbea84}3.51 &\cellcolor[HTML]{f8696b}3.42 \\\midrule
\multirow{3}{*}{DAPS} &mel loss↓ &\cellcolor[HTML]{63be7b}\textbf{1.00} &\cellcolor[HTML]{cbdc81}1.02 &\cellcolor[HTML]{ffeb84}1.02 &\cellcolor[HTML]{f8696b}1.08 &\cellcolor[HTML]{cbdc81}1.02 &\cellcolor[HTML]{ffdb81}1.03 &\cellcolor[HTML]{ffe784}1.02 &\cellcolor[HTML]{9fcf7e}1.01 &\cellcolor[HTML]{ffeb84}1.02 \\
&STFT loss↓ &\cellcolor[HTML]{f2e783}1.56 &\cellcolor[HTML]{cbdc81}1.56 &\cellcolor[HTML]{63be7b}\textbf{1.55} &\cellcolor[HTML]{f8696b}1.71 &\cellcolor[HTML]{ffe383}1.57 &\cellcolor[HTML]{ffdb81}1.58 &\cellcolor[HTML]{ffeb84}1.57 &\cellcolor[HTML]{ffeb84}1.57 &\cellcolor[HTML]{97cd7e}1.56 \\
&PESQ↑ &\cellcolor[HTML]{63be7b}\textbf{3.52} &\cellcolor[HTML]{d9e082}3.48 &\cellcolor[HTML]{fdc67c}3.44 &\cellcolor[HTML]{f8696b}3.38 &\cellcolor[HTML]{fee482}3.46 &\cellcolor[HTML]{ffeb84}3.47 &\cellcolor[HTML]{fdd780}3.46 &\cellcolor[HTML]{b3d580}3.49 &\cellcolor[HTML]{f6e984}3.47 \\\midrule
\multirow{3}{*}{Libri-TTS-R-clean} &mel loss↓ &\cellcolor[HTML]{63be7b}\textbf{0.39} &\cellcolor[HTML]{ffea84}0.40 &\cellcolor[HTML]{fdba7b}0.43 &\cellcolor[HTML]{f8696b}0.48 &\cellcolor[HTML]{ffeb84}0.40 &\cellcolor[HTML]{ffeb84}0.40 &\cellcolor[HTML]{79c47c}0.39 &\cellcolor[HTML]{e8e482}0.39 &\cellcolor[HTML]{e8e482}0.39 \\
&STFT loss↓ &\cellcolor[HTML]{63be7b}\textbf{0.59} &\cellcolor[HTML]{d8df81}0.60 &\cellcolor[HTML]{fdc17c}0.65 &\cellcolor[HTML]{f8696b}0.74 &\cellcolor[HTML]{ffe984}0.61 &\cellcolor[HTML]{ffeb84}0.60 &\cellcolor[HTML]{76c37c}0.60 &\cellcolor[HTML]{ebe582}0.60 &\cellcolor[HTML]{ffeb84}0.60 \\
&PESQ↑ &\cellcolor[HTML]{92cc7e}3.74 &\cellcolor[HTML]{fee583}3.69 &\cellcolor[HTML]{fba175}3.47 &\cellcolor[HTML]{f8696b}3.29 &\cellcolor[HTML]{fee883}3.69 &\cellcolor[HTML]{d7e082}3.71 &\cellcolor[HTML]{63be7b}\textbf{3.75} &\cellcolor[HTML]{ffeb84}3.70 &\cellcolor[HTML]{b8d780}3.72 \\\midrule
\multirow{3}{*}{Libri-TTS-R-other} &mel loss↓ &\cellcolor[HTML]{63be7b}\textbf{0.30} &\cellcolor[HTML]{ffeb84}0.30 &\cellcolor[HTML]{fdba7b}0.33 &\cellcolor[HTML]{f8696b}0.36 &\cellcolor[HTML]{ffe984}0.30 &\cellcolor[HTML]{ffeb84}0.30 &\cellcolor[HTML]{63be7b}0.30 &\cellcolor[HTML]{ffeb84}0.30 &\cellcolor[HTML]{ffe984}0.30 \\
&STFT loss↓ &\cellcolor[HTML]{63be7b}\textbf{0.45} &\cellcolor[HTML]{b1d47f}0.46 &\cellcolor[HTML]{fdc07c}0.49 &\cellcolor[HTML]{f8696b}0.55 &\cellcolor[HTML]{ffe984}0.46 &\cellcolor[HTML]{ffeb84}0.46 &\cellcolor[HTML]{8ac97d}0.46 &\cellcolor[HTML]{d8df81}0.46 &\cellcolor[HTML]{ffea84}0.46 \\
&PESQ↑ &\cellcolor[HTML]{77c47d}\textbf{3.73} &\cellcolor[HTML]{fee783}3.68 &\cellcolor[HTML]{fa9f75}3.47 &\cellcolor[HTML]{f8696b}3.31 &\cellcolor[HTML]{feea83}3.69 &\cellcolor[HTML]{e6e483}3.70 &\cellcolor[HTML]{63be7b}3.74 &\cellcolor[HTML]{ffeb84}3.69 &\cellcolor[HTML]{b0d580}3.71 \\\midrule
\multirow{3}{*}{Libri-TTS-clean} &mel loss↓ &\cellcolor[HTML]{63be7b}\textbf{0.45} &\cellcolor[HTML]{d2de81}0.45 &\cellcolor[HTML]{e8e482}0.46 &\cellcolor[HTML]{fa8a72}0.47 &\cellcolor[HTML]{ffeb84}0.46 &\cellcolor[HTML]{fdb57a}0.46 &\cellcolor[HTML]{f9746e}0.47 &\cellcolor[HTML]{d2de81}0.45 &\cellcolor[HTML]{f8696b}0.47 \\
&STFT loss↓ &\cellcolor[HTML]{63be7b}\textbf{0.68} &\cellcolor[HTML]{b9d780}0.68 &\cellcolor[HTML]{ffeb84}0.69 &\cellcolor[HTML]{f9706d}0.71 &\cellcolor[HTML]{ffeb84}0.69 &\cellcolor[HTML]{fed380}0.69 &\cellcolor[HTML]{f9766e}0.71 &\cellcolor[HTML]{b9d780}0.68 &\cellcolor[HTML]{f8696b}0.71 \\
&PESQ↑ &\cellcolor[HTML]{63be7b}\textbf{3.46} &\cellcolor[HTML]{e7e483}3.40 &\cellcolor[HTML]{fee082}3.39 &\cellcolor[HTML]{fbaf78}3.36 &\cellcolor[HTML]{f3e884}3.40 &\cellcolor[HTML]{ffeb84}3.39 &\cellcolor[HTML]{f8696b}3.32 &\cellcolor[HTML]{e7e483}3.40 &\cellcolor[HTML]{fa9673}3.34 \\\midrule
\multirow{3}{*}{Libri-TTS-other} &mel loss↓ &\cellcolor[HTML]{63be7b}\textbf{0.36} &\cellcolor[HTML]{bcd780}0.37 &\cellcolor[HTML]{a5d17e}0.36 &\cellcolor[HTML]{f9746e}0.38 &\cellcolor[HTML]{ffeb84}0.37 &\cellcolor[HTML]{fcaa78}0.37 &\cellcolor[HTML]{f9746e}0.38 &\cellcolor[HTML]{e8e482}0.37 &\cellcolor[HTML]{f8696b}0.38 \\
&STFT loss↓ &\cellcolor[HTML]{63be7b}\textbf{0.54} &\cellcolor[HTML]{a9d27f}0.55 &\cellcolor[HTML]{b8d67f}0.55 &\cellcolor[HTML]{f8696b}0.58 &\cellcolor[HTML]{ffeb84}0.55 &\cellcolor[HTML]{fecf7f}0.56 &\cellcolor[HTML]{f97a6f}0.57 &\cellcolor[HTML]{d4de81}0.55 &\cellcolor[HTML]{f9756e}0.58 \\
&PESQ↑ &\cellcolor[HTML]{63be7b}\textbf{3.16} &\cellcolor[HTML]{c3da81}3.12 &\cellcolor[HTML]{c5db81}3.12 &\cellcolor[HTML]{fcba7a}3.06 &\cellcolor[HTML]{ffeb84}3.09 &\cellcolor[HTML]{fcbc7b}3.06 &\cellcolor[HTML]{f8696b}3.00 &\cellcolor[HTML]{fbea84}3.09 &\cellcolor[HTML]{f98670}3.02 \\
\bottomrule
\end{tabular}
\end{table*}

\subsection{Test Datasets}
\label{sub:datasets}

Both in the final objective evaluation and the ablation studies we assessed the objective metrics on the following test dataset:
\begin{itemize}
\item \emph{Studio}: a proprietary set of $1024\times2$ studio-quality samples for male and female speakers, sampled at 22.05kHz. 
\item \emph{DAPS}: a held-out set of $128$ samples from the full-band high-fidelity \emph{DAPS} dataset
\item \emph{LibriTTS-R-clean}: a random set of $1024$ samples from the held-out set \emph{LibriTTS-R-test-clean}, with unseen speakers.
\item \emph{LibriTTS-R-other}: a random set of $1024$ samples from the held-out set \emph{LibriTTS-R-test-other}, with unseen speakers.
\item \emph{LibriTTS-clean}: a random set of $1024$ samples from the held-out set \emph{LibriTTS-test-other}, with unseen speakers.
\item \emph{LibriTTS-other}: a random set of $1024$ samples from the held-out set \emph{LibriTTS-test-other}, with unseen speakers.
\end{itemize}





\section{Results}

\subsection{Objective metrics}

The objective metrics for the retrained models are presented in Table~\ref{table:obj_res}. As one can notice, all the objective metrics consistently improve over all the test sets, and the improvement becomes more perceptually significant for 
smaller quantization codebooks. The original high-rate models were found virtually transparent to the recordings when perceptually assessed, so their improvement was not perceptually significant. However, when the number of codebooks is reduced, the deterioration of the original model becomes apparent and the improvement becomes more perceptually salient.

\subsection{Subjective evaluations}
\label{subsub:eval}
We selected the 4-codebook (3 kbps) and 2-codebook (1.5 kbps) models for further subjective listening test evaluation. 16 subjects participated in the MUSHRA~\cite{MUSHRA} test, assessing the 4 systems (2- and 4-codebook systems, original, and retrained) and a hidden PCM reference signal. 30 test stimuli were randomly selected from \emph{LibriTTS-r-clean}, \emph{LibriTTS-r-other}, \emph{LibriTTS-clean}, \emph{LibriTTS-other}, \emph{DAPS}, and \emph{Studio} test sets, 6 stimuli for each set. The MUSHRA average scores with $95\%$ confidence intervals are presented in Figure~\ref{fig:mushra}. Statistical significance of the difference from PCM for each system was assessed 
via the Wilcoxon Rank Sum test~\cite{wilcoxon1945}, revealing that the perceptual difference between the retrained 4-codebook system outputs and the original recordings is statistically insignificant (Table~\ref{table:wilcoxon})\footnote{Sample page is available here: 
\url{ibm.biz/IS24SpeechRVQ}}.


\subsection{Data Ablation Studies}
\label{sub:dataabl}
In a series of training-data ablation studies, we investigated the impact of 
excluding portions of the training data on the quality of speech reconstruction, as assessed on six held-out test datasets of various quality levels,
as described in Section~\ref{sub:datasets}.
The results for the retrained 3-kbps model after 200k training steps (with a mini-batch of size $B=72$) are presented in Table~\ref{tab:data_ablation}. We omitted STOI metric in the ablation table, as it had almost identical values in all the columns. The studies revealed several interesting observations. One can notice that, in general, medium to low-quality data is important for the reconstruction of most of the datasets, including the high-fidelity data (\emph{Studio}, \emph{DAPS}), although for some test sets (\emph{LibriTTS-R-clean}, \emph{LibriTTS-R-other}) it is not the case. We also observed that eliminating the high-quality \emph{LibriTTS-R} negatively impacted all scores, including those of its corresponding medium-quality counterpart.
On the other hand, the high-quality \emph{LibriTTS-R} test set did not seem to benefit from the presence of medium-quality \emph{LibriTTS} training data.




\section{Summary}
In this paper we (i) presented an improved version of the RVQGAN audio codec, specialized to speech-only data; (ii) showcased the importance of balanced speech data for indistinguishable reconstruction quality; and
(iii) provided an ablation study showing the significant impact of data selection for training.
Our pretrained model can be useful for speech synthesis, speech continuation, and various other tasks due to its lower bitrate and superior quality. While the current model underwent training and testing exclusively in English, we intend to extend its applicability to achieve consistent performance across multiple languages.

\bibliographystyle{IEEEtran}
\bibliography{mybib}

\end{document}